\documentclass[useAMS,usenatbib]{mn2e}

\usepackage{epsfig,graphicx,mathptmx}
\usepackage{times}
\usepackage{natbib}
\usepackage{amsmath}

\newcommand{\mone}{$^{-1}$}
\newcommand{\sqr}{$^2$}
\newcommand{\cub}{$^3$}
\newcommand{\bpar}{$b$-parameter}
\newcommand{\lcdm}{$\Lambda$CDM}
\newcommand{\lcdmn}{\lcdm$\nu$}
\newcommand{\muk}{$\umu$K}
\newcommand{\omegam}{$\Omega_{\rm m}$}
\newcommand{\omegacdm}{$\Omega_{\rm CDM}$}
\newcommand{\omegal}{$\Omega_\Lambda$}
\newcommand{\omegab}{$\Omega_{\rm b}$}
\newcommand{\omeganu}{$\Omega_\nu$}
\newcommand{\sigmae}{$\sigma_8$}
\newcommand{\hzero}{$H_0$}
\newcommand{\fnu}{$f_\nu$}
\newcommand{\fstar}{$f_\star$}
\newcommand{\zre}{$z_{\rm re}$}
\newcommand{\smnu}{$\Sigma m_\nu$}
\newcommand{\smnut}{$\Sigma m_\nu=(0,0.15,0.3,0.6)$ eV}
\newcommand{\smnunz}{$\Sigma m_\nu=(0.15,0.3,0.6)$ eV}
\newcommand{\gadgetiii}{\textsc{gadget-iii}}
\newcommand{\msun}{$M_{\sun}$}
\newcommand{\hmone}{$\,h^{-1}$}
\newcommand{\dksz}{$\mathcal{D}^{\rm kSZ}_{\ell}$}
\newcommand{\dkszt}{$\mathcal{D}^{\rm kSZ}_{3000}$}
\newcommand{\dksztp}{$\mathcal{D}^{\rm kSZ,patchy}_{3000}$}
\newcommand{\dtsz}{$\mathcal{D}^{\rm tSZ}_{\ell}$}
\newcommand{\wmap}{{\it WMAP}}
\newcommand{\planck}{{\it Planck}}
\DeclareMathAlphabet{\mathcal}{OMS}{cmsy}{m}{n}

\def\prd{Phys. Rev. D}
\def\aap{A\&A}
\def\apj{ApJ}

\def\apjl{ApJ}
\def\mnras{MNRAS}

\def\physrep{Phys. Rep.}

\def\apss{Ap\&SS}      
\def\apjs{ApJS}
\def\jcap{JCAP}        

\title[The kSZ effect and massive neutrinos]
{The kinematic Sunyaev--Zel'dovich effect of the large-scale structure~(I): dependence 
on neutrino mass}
\author[M. Roncarelli et al.]
{M. Roncarelli$^{1,2}$\thanks{E-mail: mauro.roncarelli@unibo.it},
F. Villaescusa-Navarro$^{3,4,5}$ and M. Baldi$^{1,2,6}$,
\\
$^1$Dipartimento di Fisica e Astronomia, Universit\`a di Bologna, viale Berti Pichat 
6/2, I-40127 Bologna, Italy \\
$^2$Istituto Nazionale di Astrofisica (INAF) -- Osservatorio Astronomico di Bologna, via 
Ranzani 1, I-40127 Bologna, Italy \\
$^3$Center for Computational Astrophysics, 160 Fifth Avenue, New York, NY 10010, USA \\
$^4$Istituto Nazionale di Astrofisica (INAF) -- Osservatorio Astronomico di Trieste, via 
Tiepolo 11, I-34131 Trieste, Italy \\
$^5$Istituto Nazionale di Fisica Nucleare (INFN) -- Sezione di Trieste, via Valerio 2, 
I-34127 Trieste, Italy \\
$^{6}$Istituto Nazionale di Fisica Nucleare (INFN) - Sezione di Bologna, viale Berti 
Pichat 6/2, I-40127 Bologna, Italy
}

\begin{document}

\date{Accepted 2017 January 18. Received 2017 January 17; in original form 2016 October 14}

\pagerange{\pageref{firstpage}--\pageref{lastpage}} \pubyear{2016}

\maketitle

\label{firstpage}

\begin{abstract}
  The study of neutrinos in astrophysics requires the combination of different 
  observational probes. The temperature anisotropies of the cosmic microwave background 
  induced via the kinematic Sunyaev-Zel'dovich (kSZ) effect may provide interesting 
  information since they are expected to receive significant contribution from 
  high-redshift plasma. We present a set of cosmological hydrodynamical simulations that 
  include a treatment of the neutrino component considering four different sum of 
  neutrino masses: \smnut. Using their outputs, we modelled the kSZ effect due to the 
  large-scale structure after the reionization by producing mock maps, then computed the 
  kSZ power spectrum and studied how it depends on \zre\ and \smnu. We also run a set of 
  four simulations to study and correct possible systematics due to resolution, finite 
  box size and astrophysics. With massless neutrinos we obtain \dkszt$=4.0$ \muk\sqr\ 
  (\zre=8.8), enough to account for all of the kSZ signal of \dkszt$=(2.9\pm1.3)$ 
  \muk\sqr\ measured with the South Pole Telescope. This translates into an upper limit 
  on the kSZ effect due to patchy reionization of \dksztp$<1.0$ \muk\sqr\ (95 per cent 
  confidence level). Massive neutrinos induce a damping of kSZ effect power of about 8, 
  12 and 40 per cent for \smnunz, respectively. We study the dependence of the kSZ signal 
  with \zre\ and the neutrino mass fraction, \fnu, and obtain 
  \dkszt$\propto$\zre$^{0.26}(1-f_\nu)^{14.3}$. Interestingly, the scaling with \fnu\ is 
  significantly shallower with respect to the equivalent thermal SZ effect, and may be 
  used to break the degeneracy with other cosmological parameters.
\end{abstract}



\section{INTRODUCTION} \label{sec:intro}

In the last decade the study of the neutrinos properties has seen an increasing interest 
in both the fields of particle physics and astrophysics. While the particle physics 
Standard Model (SM) considers the existence of three active massless neutrino species, 
namely the electron, mu and tau neutrinos, the detection of leptonic flavour oscillation 
of solar and atmospheric neutrinos \citep{fukuda98,ahmad01,ahmad02} indicates that these 
particles have a non-zero mass. The following studies in this field allowed to fix a 
lower limit on the sum of the three neutrino masses of \smnu$> 0.06$ eV\footnote{This 
limit applies in the case of normal hierarchy, for the inverted hierarchy the limit is 
\smnu$>0.1$ eV.} \citep[95 per cent confidence level (CL), see][and references therein]
{lesgourgues06,lesgourgues12,lesgourgues14,lesgourgues13,gonzalezgarcia14}.

From the astrophysical point of view, the existence of massive neutrinos influences the 
evolution of the large-scale structure (LSS) formation via gravitational interaction, 
thus requiring a generalization of the standard $\Lambda$ cold dark matter (\lcdm) 
cosmological model to account for \smnu\ as an additional free parameter, leading to a 
\lcdmn\ scenario. The effect of this new component is twofold. On early times it  
contributes to the Universe energy budget as radiation, with density of
\begin{equation}
\rho_\nu=\frac{7}{8}\left(\frac{4}{11}\right)^\frac{4}{3}N_{\rm eff}\,\rho_\gamma\, ,
\end{equation}
where $\rho_\gamma$ is the photon energy density, and $N_{\rm eff}$ is the effective 
number of neutrino species ($N_{\rm eff}=3.046$, according to the SM). This causes a 
postponing of the matter--radiation equality for a fixed value of \omegam$\,h^2$ (where 
\omegam\ is the ratio between the total matter density of the Universe and the critical 
one, $\rho_{\rm c}$, at the present epoch and $h$ is the Hubble constant $H_0$ in units 
of 100 km s\mone Mpc\mone). At late times the neutrinos large thermal velocities prevent 
their clustering on scales smaller than their free-streaming length
\begin{equation}
\lambda_{\rm fs,\nu} \simeq 
7.67\,\frac{H_0(1+z)^2}{H(z)}\left(\frac{1\,\rm eV}{m_\nu}\right) h^{-1}\,\rm Mpc.
\label{e:l_fs}
\end{equation}
This induces a gravitational backreaction that suppresses density fluctuations 
of both baryons and CDM on the same scales. Thus, as a net effect massive neutrinos cause 
a suppression in the amplitude of the matter power spectrum on small scales with respect 
to an equivalent massless neutrinos model.

The dependence of these effects on the total neutrino mass indicates that astrophysical 
observations have the potential to provide constraints on the value of \smnu, as it has 
been shown by several recent theoretical works 
\citep[see e.~g.][]{viel10,marulli11,shimon12,carbone13,costanzi13a,mak13,roncarelli15,
Palanque_2015}. In fact, the list of cosmological probes that can be used to place limits 
on the sum of neutrino masses is nowadays long and heterogeneous, including galaxy 
redshift surveys \citep{elgaroy02,tegmark06,thomas10}, galaxy clustering 
\citep{Saito_2011,zhao13,beutler14,sanchez14}, cosmic microwave background (CMB) 
observations from the {\it Wilkinson Microwave Anisotropy Probe} \citep[\wmap;][]
{komatsu09,komatsu11,hinshaw13} and \planck\ \citep{planck16cp}, Ly-$\alpha$ forest 
studies \citep{croft99,viel10,Palanque_2015}, galaxy clusters mass function 
\citep{mantz10b,mantz15} and future 21cm intensity mapping observations \citep{Paco_15b}. 

Most notably, the current tightest constraints on the sum of the neutrino masses 
have been derived by \cite{Palanque_2015} who combined data from the CMB with BAO and 
Ly-$\alpha$ forest and obtained an upper limit (95 per cent C.~L.) of \smnu$\le0.14$ eV 
\citep[see also][for a different analysis combing CMB and galaxy clustering]
{Cuesta_2016}. It should be noted, however, that all these bounds have been derived 
assuming that massive neutrinos evolve in an otherwise standard \lcdm\ cosmological 
model. Possible alternative scenarios such as non-standard dark energy or modified 
gravity theories may result in significantly looser constraints 
\citep[see e.~g.][]{LaVacca_etal_2009,He_2013,Motohashi_etal_2013,Baldi_etal_2014}. 
Furthermore, the first release of \planck\ satellite 
data highlighted a discrepancy between the cosmological parameters obtained with primary 
CMB observations \citep{planck14cp} and the relatively low number counts of galaxy 
clusters detected via thermal Sunyaev-Zel'dovich \citep[tSZ;][]{sunyaev70} effect: the 
joint analysis of these datasets and BAO data led to an estimate of \smnu$=(0.20\pm0.09)$ 
eV \citep{planck14cc}. However this result is hampered by the degeneracy with the other 
cosmological parameters (mainly \sigmae\ and \omegam) and by the uncertainty in the 
amount of non-thermal pressure contribution in galaxy clusters. Indeed, more recently 
the analysis of the latest \planck\ data release \citep{planck16cc} showed that the 
attempt to reconcile primary CMB and cluster counts by increasing \smnu\ only 
leads to a conflict with $H_0$ measurements from BAO \citep[see also the discussion in][]
{Leistedt_Peiris_Verde_2014,planck16cp}. This indicates that the origin of this tension 
is not fully understood and that additional efforts involving other observational probes 
are necessary.

In this framework new potentially interesting observational constraints may come from 
the kinematic Sunyaev-Zel'dovich (kSZ) effect, i.~e. a secondary anisotropy of the CMB 
caused by the motion of free electrons of the LSS, whose dependence on the physical 
properties of the intergalactic medium (IGM) is complementary in many aspects with 
respect to the tSZ one \citep[see e.~g.][]{roncarelli07,battaglia10,shaw12}. In 
particular, the kSZ effect is expected to be sensitive to the IGM residing in 
non-collapsed structures at high redshift. While several new generation microwave 
telescopes have reached the sensitivity to achieve the first kSZ effect detections on 
galaxy clusters \citep{hand12,sayers13,planck16ksz}, up to now the only measurement of 
its power spectrum has been obtained with the South Pole Telescope (SPT) by 
\cite{crawford14} and by \cite{george15}. By combining the tSZ bispectrum from the SPT-SZ 
survey (800 deg\sqr) with the full 2540 deg\sqr\ field they measured an amplitude of the 
kSZ temperature fluctuations of \dkszt$=(2.9\pm1.3)$ \muk\sqr\ at $\ell=3000$. This 
places the first interesting constraint on a quantity that is expected to receive 
contribution from the LSS of the ionized Universe and from the epoch of reionization 
(EoR) itself. While the latter is extremely uncertain and depends on subtle details of 
the reionization process \citep[see][and references therein]{iliev14}, the kSZ effect of 
the post-ionization era can be studied with suitable LSS simulations and depends mostly 
on the cosmological assumptions. This gives the kSZ effect the potential to provide 
constraints on several cosmological parameters, including \smnu.

This paper is the first of a series of studies of the kSZ effect from the LSS of the 
Universe in different cosmological scenarios beyond the standard \lcdm\ paradigm. In this 
work we present the first analysis of the kSZ effect derived from a set of cosmological 
hydrodynamical simulations that include a detailed treatment of the massive neutrino 
component \citep{viel10}. Our simulations follow the evolution of structure formation 
from large scales ($L_{\rm box}=240$\hmone\ Mpc) down to the non-linear regime spanning 
all the relevant redshifts. By analysing the outputs of the baryonic component, adopting 
a minimum number of assumptions, we derive a set of Doppler \bpar\ maps and compute the 
corresponding power spectrum of kSZ-induced CMB temperature fluctuations in the multipole 
range of $1000<\ell<20000$. This allows us to make a direct comparison with the SPT 
measurement and to derive how our results depend on the value of the redshift of 
reionization, \zre, and \smnu.

This paper is organized as follows. In the next section, we describe the simulation set 
used in this work, our procedure to create mock light-cones and our kSZ effect model. 
In Section~\ref{s:res} we present our results in terms of the effect of neutrino mass on 
the global properties of the kSZ effect and discuss the dependence with \zre. 
Section~\ref{s:sys} is devoted to the study of possible systematics caused by limited box 
size, finite resolution and accuracy of the astrophysical processes of our simulations. 
We summarize and draw our conclusions in Section~\ref{s:concl}.


\section{MODELLING THE EFFECT OF NEUTRINOS ON THE kSZ SIGNAL} \label{sec:models}

\subsection{The simulation set} 
\label{ss:sims}

We have run hydrodynamical simulations using the \textsc{treepm}+SPH code \gadgetiii\  
\citep{springel05} for four different cosmologies with massless and massive neutrinos. 
The value of the following cosmological parameters are common to all models: 
\omegam$=$\omegacdm$+$\omegab$+$\omeganu$=0.3175$, \omegab$=0.049$, \omegal$=0.6825$, 
$h=0.6711$, $n_s=0.9624$, $A_s=2.13\times10^{-9}$ (corresponding to \sigmae$=0.834$ at 
$z=0$ in the \lcdm\ scenario), in agreement with the latest results by \cite{planck16cp}. 
In models with massive neutrinos we set \omeganu$h^2=$\smnu$/(93.14~{\rm eV})$ and 
modified the value of \omegacdm\ accordingly to keep \omegam\ constant. In our fiducial 
runs we follow the evolution of $512^3$ CDM plus $512^3$ gas plus $512^3$ neutrino 
particles (only in models with massive neutrinos). A summary of the simulation parameters 
is shown in Table~\ref{t:sim}.

\begin{table*}
\begin{center}
\caption{
  Parameter values of our simulation suite. First column: simulation name. Second column: 
  comoving box size, in \hmone\ Mpc. Third column: total number of particles, including 
  all three species (two in case of massless neutrinos). Fourth to sixth column: masses 
  of the CDM, baryonic and neutrino particles, respectively, in units of 
  $10^{9}$\hmone\msun. Seventh to ninth column: sum of neutrino masses, \smnu, in eV, 
  corresponding value of neutrino mass fraction, \fnu=\omeganu$/$\omegam, in percent 
  units, and \sigmae\ at present epoch, respectively. The last column shows the physical 
  model(s) assumed for the baryonic component (radiative cooling and UV background are 
  present in all simulations, see text for details). All simulations assume a flat 
  \lcdm($\nu$) cosmology with \omegal$=0.6825$, 
  \omegam$=$\omegacdm$+$\omegab$+$\omeganu$=0.3175$, \hzero$=67.11$ km/s/Mpc 
  ($h=0.6711$), $n_s=0.9624$ and $A_s=2.13\times10^{-9}$. 
  }
\begin{tabular}{rcccccccccccc}
\hline
\hline
Simulation && 
$L_{\rm box}$ & $N_p$ & $m_{p, \rm cdm}$ & $m_{p,\rm b}$ & 
$m_{p,\nu}$ && \smnu\ & 10\sqr\fnu\ & \sigmae & Baryon physics \\
 &&
(\hmone\ Mpc) &             & \multicolumn{3}{c}{($10^{9}$\hmone\msun)} &&
(eV) &  & ($z=0$) & (+ rad. cooling + UV bkg.) \\
\hline
  N0 && 240 & 2$\times$  512\cub & 7.68 & 1.40 & $-$  && 0    & 0    & 0.834 & quick Ly-$\alpha$ \\
 N15 && 240 & 3$\times$  512\cub & 7.57 & 1.40 & 0.10 && 0.15 & 1.11 & 0.801 & quick Ly-$\alpha$ \\
 N30 && 240 & 3$\times$  512\cub & 7.47 & 1.40 & 0.20 && 0.30 & 2.23 & 0.764 & quick Ly-$\alpha$ \\
\vspace{0.12cm}
 N60 && 240 & 3$\times$  512\cub & 7.27 & 1.40 & 0.40 && 0.60 & 4.46 & 0.693 & quick Ly-$\alpha$ \\
  HR && 240 & 2$\times$ 1024\cub & 0.96 & 0.18 & $-$  && 0    & 0    & 0.834 & quick Ly-$\alpha$ \\
  SB && 120 & 2$\times$  256\cub & 7.68 & 1.40 & $-$  && 0    & 0    & 0.834 & quick Ly-$\alpha$ \\
  LB && 480 & 2$\times$ 1024\cub & 7.68 & 1.40 & $-$  && 0    & 0    & 0.834 & quick Ly-$\alpha$ \\
CSFW && 240 & 2$\times$  512\cub & 7.68 & 1.40 & $-$  && 0    & 0    & 0.834 & star form. + SN feedback\\
\hline
\hline
\label{t:sim}
\end{tabular}
\end{center}
\end{table*}

The initial conditions of the simulations have been generated by placing each particle 
species (CDM, gas and neutrinos) in a different regular cubic grid, then displacing and 
assigning them peculiar velocities using the Zel'dovich approximation. To this purpose 
we rescaled the matter $z=0$ power spectra and transfer functions, computed with 
\textsc{camb} \citep{CAMB}, to the starting redshift of the simulations that we fix to 
$z=99$. This rescaling was done using the Newtonian physics implemented in the numerical 
simulation. For cosmologies with massive neutrinos this is translated into a 
scale-dependence growth factor and rate inboth the CDM and neutrino power spectra
\footnote{The rescaling is performed using the \textsc{reps} code: 
https://github.com/matteozennaro/reps.}. For further details we refer the reader to 
\cite{Zennaro_2016}. We also assigned thermal velocities, besides peculiar ones, to 
neutrino particles by randomly sampling the neutrino Fermi-Dirac momentum distribution. 

In our first simulation set, dubbed N0, N15, N30 and N60 in Table~\ref{t:sim}, we fixed 
the comoving box size to $L_{\rm box} = 240$\hmone\ Mpc and varied only the neutrino mass 
assuming four different values: \smnut, respectively. All other parameters have been kept 
identical, including the random seeds of the initial conditions, ensuring that we 
actually model the evolution of the same structures down to $z=0$. In order to study 
systematic effects (see Section~\ref{s:sys}) we have run a second set of four simulations 
assuming massless neutrinos. The first one, dubbed HR, has higher resolution 
(2$\times$1024\cub\ particles) and same box size. The following two, SB and LB, have 
different comoving box sizes, (120, 480) \hmone\ Mpc, respectively, to investigate the 
dependence of our results on the simulation volume: in order to keep the same mass 
resolution, we used 2$\times$256\cub\ and 2$\times$1024\cub\ particles for the SB and LB 
runs, respectively. Finally, a last simulation, dubbed CSFW, assumes same resolution and 
box size of the N0 but adopts a different physical treatment for the baryonic component, 
allowing us to study the dependence of our results with the astrophysical processes that 
affect the IGM. For each simulation we have produced a set of 26 snapshots at different 
redshifts from $z=15$ down to the present epoch. This ensures that we can reconstruct the 
full line of sight of the kSZ signal.

All our hydrodynamical simulations include radiative cooling by hydrogen and helium and 
heating by a uniform UV background. The cooling routine has been modified in order to 
reproduce the measured Universe thermal history as in \cite{Viel_13}. In order to save 
computational time we have run our simulations (except the CSFW simulation) using the 
so-called `quick Ly-$\alpha$' technique. This method consists in following the 
hydrodynamical evolution of the gas particles until their densities reach a given 
threshold, beyond which the code turns them into star particles and turns off the 
hydrodynamic forces. In order to verify that our results are not largely affected by
this simplistic physics model, in our CSFW run we implemented star formation and 
supernova (SN) feedback following \cite{SH_2003}, instead of the quick Ly-$\alpha$ 
method. SN feedback is implemented as kinetic feedback, with constant velocity winds of 
350 km/s.

\begin{figure}
\includegraphics[width=0.5\textwidth]{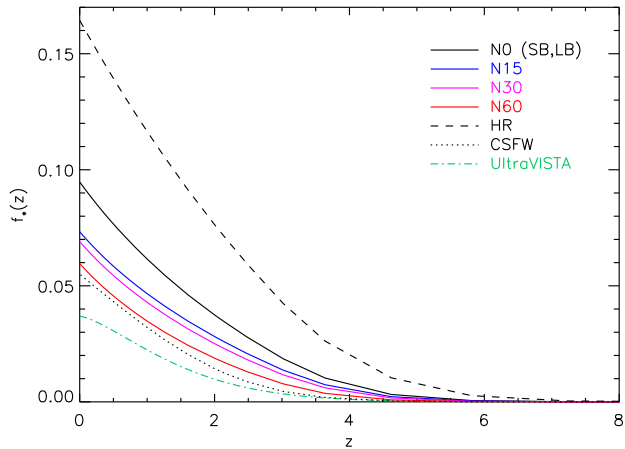}
\caption{
  Global star mass fraction ($\rho_\star / \rho_{\rm b}$) in the whole box volume as a 
  function of redshift for the eight simulations. Solid lines indicate our first 
  simulation set with \smnut\ shown in black, blue, magenta and red, respectively. The 
  black dashed and dotted lines correspond to the HR and CSFW simulations, respectively. 
  The values of \fstar$(z)$ for the SB and LB simulations (not shown) are very close to 
  the N0 ones (see details in Table~\ref{t:sim}). The green dot--dashed line is derived 
  from the cosmic stellar mass density estimated by \protect\cite{ilbert13}.
  }
\label{f:fstar}
\end{figure}

While the modelling of the X-ray emission and of the tSZ effect depends highly on the 
assumptions on cooling and feedback, the global kSZ signal due to the LSS after the 
reionization shows a much weaker dependence on the physical scheme adopted for the 
baryonic component\footnote{This, of course, is not true for the possible patchy kSZ 
effect imprinted during the EoR, which would add to the one subject of our study.}, with 
the only meaningful uncertainty that resides in the amount of cosmic gas that is not in 
the ionized state \citep[see e.~g.][]{roncarelli07,trac11,shaw12}. Given the negligible 
amount of mass associated with neutral and molecular hydrogen, this translates into 
determining the correct stellar mass fraction as a function of redshift. It is well known 
that the quick Ly-$\alpha$ cooling scheme can not reproduce a realistic star-formation 
history, as it can be seen in the plot of Fig.~\ref{f:fstar} that shows the stellar mass 
fraction, \fstar$\equiv\rho_\star/\rho_{\rm b}$, as a function of redshift obtained from 
the raw output of our simulations. As expected, the main difference is due to the change 
in the gas particle resolution, with higher resolution predicting higher star formation, 
and with neutrino mass changes that produce smaller effects. We compare these results to 
the ones obtained with the UltraVISTA data by \citeauthor{ilbert13} (\citeyear{ilbert13}, 
green dot-dashed line): we observe that in all cases our models predict a significant 
excess of star formation, with our N0 model that ends at $z=0$ with \fstar$\simeq 9.5$ 
per cent compared to the observed \fstar$\simeq 3.8$ per cent. The CSFW model, thanks to 
its feedback scheme, shows more realistic values of \fstar but still in excess with 
respect to the observed ones. To overcome this issue, for the purpose of this work we 
assume the \fstar$(z)$\ value by \cite{ilbert13} and apply a correction a-posteriori to 
the differential kSZ signal at all redshifts to account for this difference. We will 
discuss this point in detail in Section~\ref{ss:ksz}.

\subsection{Light-cone construction} 
\label{ss:lcone}

In order to model the kSZ effect integrated along the line of sight we need to use the 
outputs of our simulations to reconstruct the full light-cone geometry from $z=0$ up to 
the EoR. Our method is similar to the one adopted in our previous papers \citep[see 
e.~.g.][]{roncarelli10a,roncarelli12,roncarelli15}: we refer the reader to these works 
for the details, and we just summarize here the main points.

We stack the different simulation volumes along the line of sight up to $z=15$, 
corresponding to 7022\hmone\ comoving Mpc. To avoid the repetition of the same 
structures along the line of sight, we randomize every simulation volume in the following 
ways: (i) we assign to each Cartesian axis a 50 per cent probability to be reflected, 
(ii) we randomly pick the axis oriented along the line of sight and the two perpendicular 
ones, (iii) we recenter randomly the coordinates by using the periodic boundary 
conditions and, (iv) for distances that assure that the box volume does not exceed the 
opening angle of the light-cone, the cube is rotated by a random angle around the line of 
sight passing through its centre. This process is done using the same random seeds for 
all simulations: since we are also using the same phases in the initial conditions, this 
ensures that we represent exactly an identical realization of the comoving 
volume\footnote{This does not apply completely for the SB and LB simulations, whose 
comoving box has a different size.} with different models, eliminating the effect of 
cosmic variance when comparing them. To increase the statistical accuracy of our results, 
we generate 50 different light-cone realizations by varying the seed of the full 
randomization process.

In order to optimize the redshift sampling, each simulation volume is divided along the 
line of sight into slices associated with different simulation outputs: their limits are 
defined in a way that ensures that for every comoving distance we chose the snapshot that 
better matches the corresponding age of the Universe in our cosmological model.

The maximum opening angle of the light-cone is limited by the angular box size at $z=15$: 
this allows us to fix it to 1.8$^\circ$ for all simulations, except the SB for which we 
are limited to 0.9$^\circ$. When accounting for the 50 light-cone realizations this 
corresponds to a total sky area of 162 deg\sqr\ (40.5 deg\sqr\ for the SB). We point out 
that this area can be considered fully independent only up to $z\sim 0.4$ because at 
higher redshifts the light-cones geometry starts to replicate to some extent the same 
structures in the different realizations, and includes almost all the simulation volume 
at $z=15$. However, this issue is compensated by the decrease of cosmic variance at high 
redshift.

\begin{figure*}
\vspace{-5cm}
\includegraphics{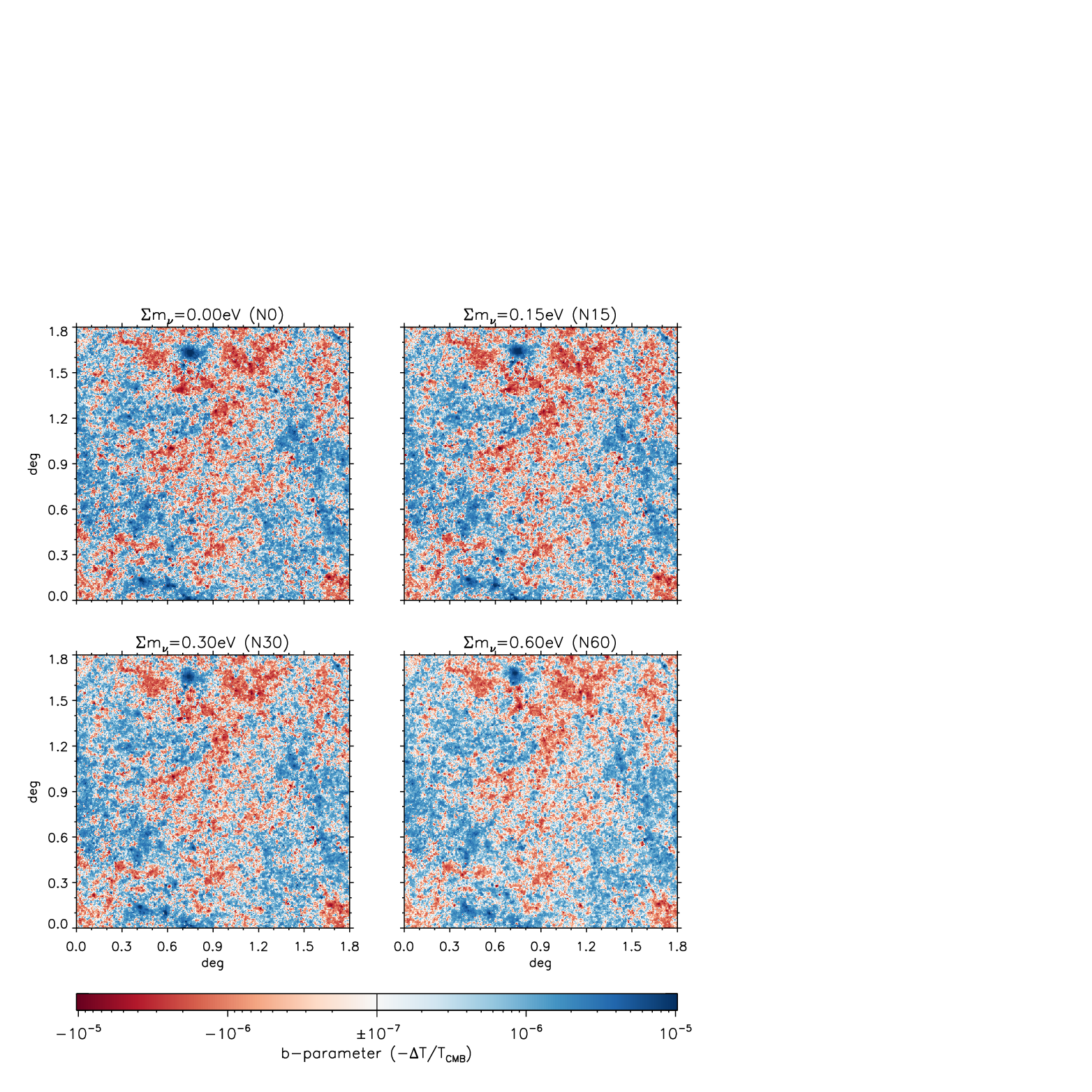}
\caption{
  Maps of the \bpar\ as a function of \smnu. Each map is 1.8$^\circ$ per side with a 
  resolution of 6.33 arcsec ($1024^2$ pixels) and represents the signal integrated 
  from $z=0$ to \zre$=8.8$ for the same light-cone assuming massless neutrinos 
  (top-left) and \smnunz\ (top-right, bottom-left and bottom-right, respectively). 
  The colour scale indicates in red the sky regions where, on average, the gas is 
  approaching the observer ($b<0$ and increase of observed CMB temperature), and in blue 
  where the gas is receding ($b>0$ and decrease of observed CMB temperature).
  }
\label{f:maps}
\end{figure*}

\subsection{The kSZ effect model} 
\label{ss:ksz}

The kSZ effect \citep[also referred to as the Ostriker-Vishniac effect, in the 
linear-theory approximation,][]{ostriker86,vishniac87} is the Doppler shift of the CMB 
photons caused by the interaction with free electrons of the IGM that have a proper 
motion with respect to the CMB rest frame. Its intensity in a given direction 
with unit vector $\hat{\gamma}$ can be expressed using the Doppler \bpar, defined as:
\begin{equation}
b(\hat \gamma) \equiv 
\frac{\sigma_{\rm T}}{c} \int_0^{z_{\rm re}} \! \frac{{\rm d}\chi}{{\rm d}z} \frac{{\rm d}z}{1+z} \, 
e^{-\tau(z)} \, n_{\rm e} \, \vec v_{\rm e} \cdot \hat \gamma \, ,
\label{e:bdef}
\end{equation}
where $\sigma_{\rm T}$ is the Thomson cross-section, $c$ is the light speed in vacuum, 
$n_{\rm e}$ is the electron number density and $\vec v_{\rm e}$ is the proper velocity of the 
electrons. The integral is calculated from the observer up to \zre\ along the comoving 
coordinate $\chi$. In the previous equation $\tau$ is the Thomson optical depth:
\begin{equation}
\tau(z) \equiv \sigma_{\rm T}\int_0^{z}\!\frac{{\rm d}\chi}{{\rm d}z'}\frac{{\rm d}z'}{1+z'}\,n_{\rm e}(z') \, .
\label{e:tau}
\end{equation}

Once the value of $b$ is known, the corresponding change in the observed CMB temperature 
measurement is simply
\begin{equation}
\Delta T = -b \, T_{\rm CMB} \, ,
\label{e:dt}
\end{equation}
thus independent of observational frequency, with approaching (receding) gas that induces 
a temperature increment (decrement).

While it is known that possible non-uniform reionization scenarios can imprint a 
non-negligible kSZ effect associated with the EoR \citep[see e.~g.][and references 
therein]{mcquinn05,zahn05,iliev07,iliev08,mesinger12}, this would require simulations 
that include a detailed modelling of both reionization sources and radiative transfer, 
and is therefore beyond the scope of this paper. Since we focus only on the kSZ of the 
LSS after the EoR, we can safely assume that the Universe is fully ionized, leaving \zre\ 
as a free parameter, and apply instead the following equation:
\begin{equation}
b(\hat \gamma) = \frac{\sigma_{\rm T}X x_{\rm e}}{c \, m_{\rm p}} \int_0^{z_{\rm re}} \! 
                 \frac{{\rm d}\chi}{{\rm d}z} \frac{{\rm d}z}{1+z} \, e^{-\tau(z)} \, \rho_{\rm g} \, 
                 \vec v_{\rm g} \cdot \hat \gamma \, ,
\label{e:bapp}
\end{equation}
being $X=0.76$ the cosmological hydrogen mass fraction, $x_{\rm e} \simeq 1.16$ the 
electron-to-proton ratio in a fully ionized primordial plasma, $m_{\rm p}$ the proton mass and 
$\rho_{\rm g}$ the gas mass density.

As said in Section~\ref{ss:sims}, with this approach the main uncertainty resides 
in an accurate modelling of the star mass fraction, since one can safely assume that 
$\rho_{\rm g}=\rho_{\rm b} (1-f_\star)$. The output of our hydrodynamical 
simulations already provides the value of the cosmological gas density, however, given 
the bias shown in Fig.~\ref{f:fstar}, we correct the output in the following way:
\begin{equation}
\rho_{\rm g} (z) = 
\rho_{\rm g,sim}(z)\left[\frac{1-f_{\star,\rm obs}(z)}{1-f_{\star,\rm sim}(z)}\right]\, ,
\label{e:rhoc}
\end{equation}
where $f_{\star, \rm sim}(z)$ and $f_{\star, \rm obs}(z)$ are the star mass fraction from 
the simulation and from \cite{ilbert13} results, respectively, as a function of redshift. 
This correction is applied globally to all the signal coming from the same redshift slice 
(see Section~\ref{ss:lcone}), thus neglecting the actual spatial distribution of star 
particles. However, this modification produces only minor differences in our final 
results (see the discussion in Section~\ref{s:sys}).

\subsection{The mapping procedure} 
\label{ss:maps}

We used our simulations to create a set of \bpar\ maps from which we derive our results 
on the kSZ effect. The mapping procedure is coherent with the SPH formalism and is 
described in detail in our previous works \citep[see][]{roncarelli06a,roncarelli07,
ursino10,roncarelli12}. We summarize here the main points.

For every SPH particle that falls inside the light-cone we consider the output physical 
quantities that determine the kSZ effect, namely its mass, $m_i$, radial velocity, 
$v_{r,i}\equiv \vec v_i \cdot \hat \gamma$, and angular diameter distance from the 
observer, $d_{A,i}$. We also include particles whose centre falls outside the light-cone 
but whose sphere with radius given by its SPH smoothing length, $h_i$, intersects 
its border. We then determine the integrated \bpar\ of each SPH particle as
\begin{equation}
B_i = 
\frac{X\,\sigma_{\rm T}\,x_{\rm e}}{m_{\rm p}\,c\,d_{A,i}^2} \, 
e^{-\tau(z_i)}\,m_i\,v_{r,i} \, ,
\label{e:bsph}
\end{equation}
where $z_i$ is the cosmological redshift in the particle's position. Finally, the values 
of $B_i$ are used to convert the line of sight integral of equation~(\ref{e:bapp}) into a 
sum over the SPH particles and distributed over the map pixels close to the particle's 
centre, by adopting the same smoothing kernel used by the hydrodynamical code.

For the purpose of our work we create maps of 1024 pixels per side. Given the map size of 
1.8$^\circ$ per side (0.9$^\circ$ for the SB simulation), this corresponds to an angular 
resolution of 6.33 arcsec (3.16 arcsec). This procedure is repeated for all the 
simulations and for the 50 light-cones. In addition, to study the redshift dependence of 
our results and to account for the uncertainty in the value of \zre, we compute the 
integral separately for 20 logarithmically equi spaced redshift bins. This makes a total 
of 1000 \bpar\ maps for each simulation.


\section{RESULTS} \label{s:res}

\subsection{Global properties of the Doppler \bpar} 
\label{ss:glob}

We show in Fig.~\ref{f:maps} the \bpar\ maps obtained with our method for different 
values of \smnu\ (first simulation set) for the same light-cone realization. For these 
maps we assumed as integration limit (see equations~\ref{e:bdef} and \ref{e:bapp}) the 
nominal value of \zre$=8.8$ from \cite{planck16cp}. The typical values are of the order 
of $|b| \approx 10^{-6}$, with peaks that reach $10^{-5}$ associated with galaxy clusters 
and groups. However, the main coherent structures that can be seen in these maps are at 
larger angular scales, about 10 arcmin, and are associated with the bulk motions of the 
LSS. As expected, the presence of massive neutrinos slows down the growth of structure 
formation, damping velocity amplitudes at all redshifts. The consequence is that for 
increasing neutrino mass the amplitude of the kSZ effect is significantly reduced, both 
for peaks and for typical values: this can be seen by comparing qualitatively the four 
maps, where the same structures are present, with lower intensities for increasing \smnu.

\begin{figure}
\includegraphics[width=0.5\textwidth, trim=0.5cm 0 -0.5cm 0]{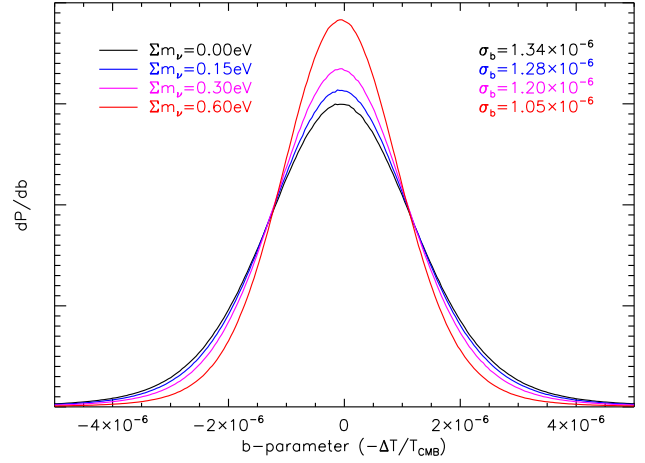}
\caption{
  Probability distribution function of the \bpar\ for different neutrino masses. Each 
  curve is computed for the whole set of 50 light cones (i.~e. 50 sets of maps like the 
  ones shown in Fig.~\ref{f:maps}) with signal integrated up to \zre$=8.8$ and pixels of 
  6.33 arcsec per side. The black line indicates the model with massless neutrinos while 
  \smnu$=(0.15,0.3,0.6)$ eV are shown in blue, magenta and red, respectively. On the 
  right hand side we show the standard deviation of the best-fitting Gaussian 
  distribution of each curve.
  }
\label{f:bdist}
\end{figure}

This effect is shown in detail in Fig.~\ref{f:bdist}, where we plot the probability 
distribution functions of the map pixels, computed considering the full set of 50 
light-cones. All distributions peak around zero as expected by definition, but the 
dispersion decreases for higher values of \smnu. We find that all curves can be fit by a 
Gaussian distribution with mean and dispersion left as a free parameter. The results of 
the fit for the dispersion are quoted in the same plot, with $\sigma_b$ going 
from $1.34 \times 10^{-6}$ for massless neutrinos \citep[in agreement with][]
{roncarelli07} to $1.05 \times 10^{-6}$ in the case of \smnu$=0.6$ eV. Following 
\cite{roncarelli15}, we study its dependence on \fnu. We find that the relation between 
the two quantities can be fitted with a power-law function, and obtain that $\sigma_b$ 
scales as
\begin{equation}
\sigma_{b,\nu} = \sigma_{b,0} \, (1 - f_\nu)^{5.2} \, ,
\label{e:sbsc}
\end{equation}
being $\sigma_{b,0}$ the value for the \lcdm\ model.

\subsection{The kSZ effect power spectrum} 
\label{ss:pows}

\begin{figure}
\includegraphics[width=0.5\textwidth]{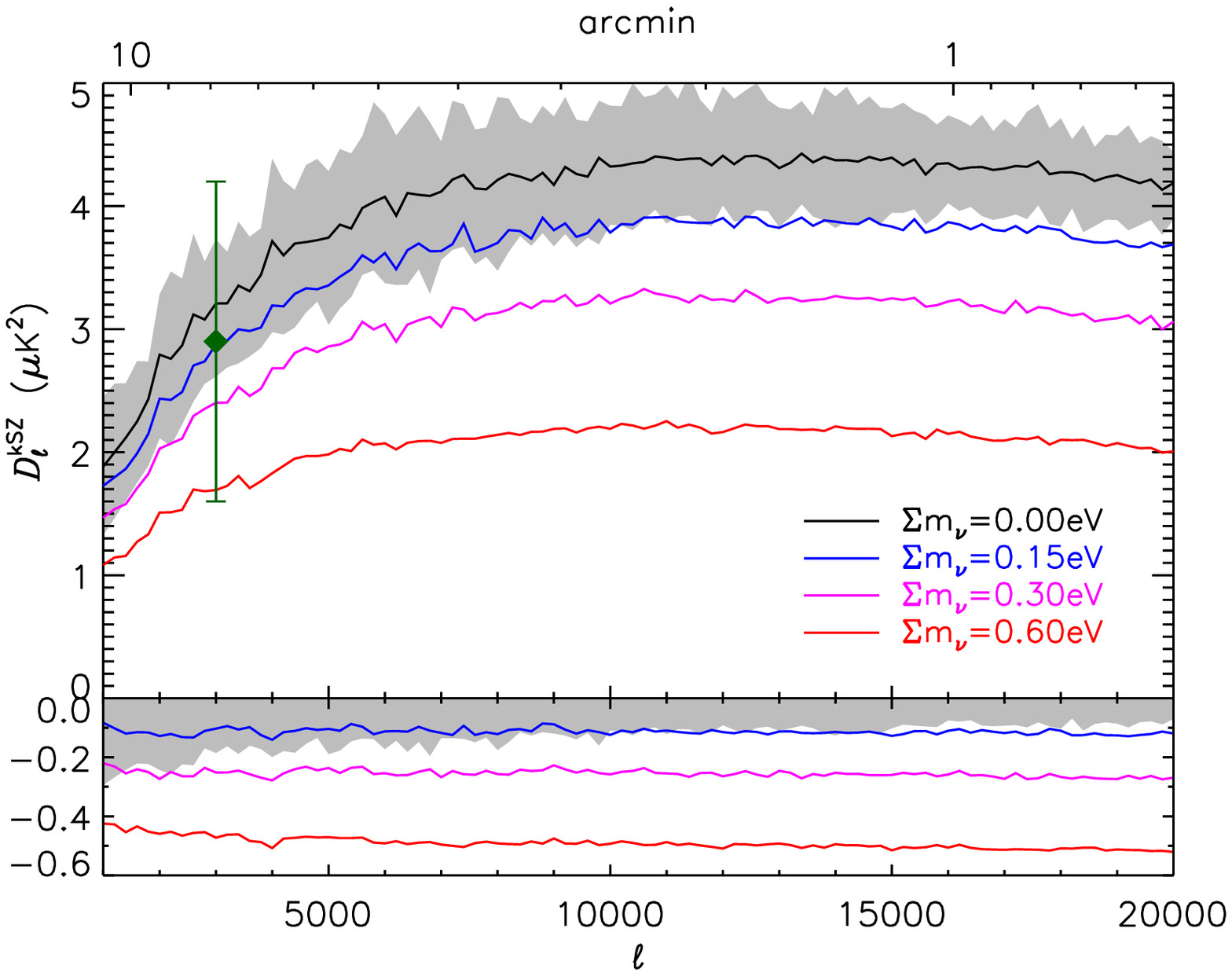}
\caption{
  Top panel: angular power spectrum of temperature fluctuations associated with the kSZ 
  effect (\dksz) as a function of the multipole $\ell$ for different values of \smnu. 
  The curves are computed by averaging over the full set of 50 light-cones (a total of 
  162 deg\sqr) and consider the signal integrated up to \zre$=8.8$. The colour coding of 
  the lines is the same as in Fig.~\ref{f:bdist}. The grey-shaded area encloses the 
  central 34 light-cones (68 per cent of the total) for the N0 model only. The green 
  diamond with error bars shows the measurement of \dkszt$=(2.9\pm1.3)$ \muk\sqr\ by 
  \protect\cite{george15}. Bottom panel: relative differences with respect to the N0 
  model.
  }
\label{f:pows}
\end{figure}

One of the most important observables for the kSZ effect is its angular power spectrum, 
whose specific features can allow to break the degeneracy with other sources of CMB 
anisotropies \citep[see e.~g.][]{crawford14}. Using our set of \bpar\ maps we first 
convert them into $\Delta T$ using equation~(\ref{e:dt}), then apply the Fast Fourier 
Transform method in the flat-sky approximation to compute the amplitude of the angular 
power spectrum of the kSZ temperature fluctuations as a function of the multipole $\ell$. 
We report the results in terms of $\mathcal{D}_{\ell}$, defined as
\begin{equation}
\mathcal{D}_{\ell} \equiv \frac{\ell(\ell+1)\,C_{\ell}}{2\upi} \, ,
\end{equation}
being $C_{\ell}$ the usual definition of power spectrum of temperature fluctuations. 
We show in Fig.~\ref{f:pows} the power spectra of our first simulation set obtained as 
the average over the 50 light-cones. In all cases the amplitude, \dksz, of the kSZ 
power spectrum increases from $\ell=1000$ to $6000$, with a flattening for higher 
multipoles: this is consistent with what was found in previous works \citep{roncarelli07,
trac11,shaw12}. The cosmic variance in the map fields for the N0 simulation 
(grey-shaded area) decreases with increasing multipole, from about 1 \muk\sqr\ at 
$\ell=1000$ to 0.5 \muk\sqr\ at $\ell=20000$. Interestingly, our N0 model shows a value 
of \dkszt$=3.21$ \muk\sqr, enough to explain all of the SPT measurement of 
\dkszt$=(2.9\pm1.3)$ \muk\sqr\ measured by \cite{george15}. Since our method 
is expected to underestimate the true value by about 20 per cent due to the limited box 
size (see Section~\ref{s:sys} for the details), this significantly reduces the amount of 
signal left that may be due to patchy reionization in a \lcdm\ cosmology consistent with 
the results of \cite{planck16cp}, as we will discuss in Section~\ref{s:sys}.

Massive neutrinos induce a suppression of the amplitude of the kSZ power spectrum 
at all multipoles, slightly more pronounced at small scales. In our most extreme model, 
N60, the power is reduced by 40 percent at $\ell=1000$ and by half at $\ell > 10000$. 
However, when assuming more realistic scenarios given by the N15 and N30 simulations, the 
decrease of power at the same scales is significantly smaller, 8 and 12 per cent, 
respectively. These differences can be compared with the kSZ measurement by 
\citeauthor{george15} (\citeyear{george15}, green diamond): all four models fall inside 
the 1$\sigma$ interval, indicating that such difference would be difficult to measure 
with current instruments. Nonetheless we point out that if a significant additional kSZ 
effect associated with the EoR is present (see e.~g. the relatively high values predicted 
by \citealp{iliev08} and \citealp{mesinger12}), then the presence of massive neutrinos 
would help to alleviate the tension between observations and theory.

\begin{figure}
\includegraphics[width=0.5\textwidth]{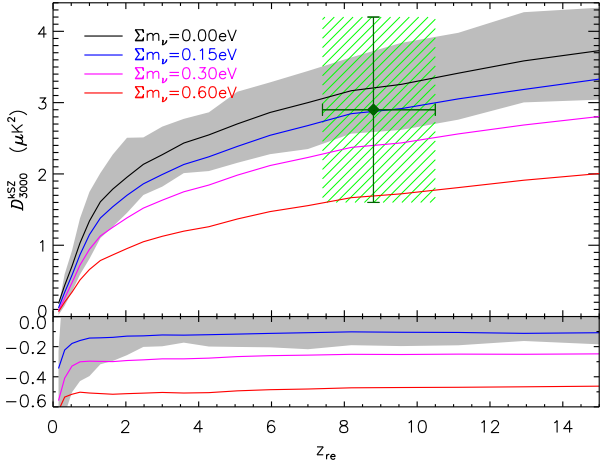}
\caption{
  Top panel: amplitude of the kSZ power spectrum at $\ell=3000$ (\dkszt) as a function of 
  \zre\ for different values of \smnu. The colour coding is the same as in 
  Fig.~\ref{f:bdist}. The grey-shaded area encloses the central 34 light-cones (68 per 
  cent of the total) for the N0 model only. The green diamond with error bars (1$\sigma$) 
  and green-shaded area shows the results of \zre$=8.8^{+1.7}_{-1.4}$ and 
  \dkszt$=(2.9\pm1.3)$ \muk\sqr\ by \protect\cite{planck16cp} and 
  \protect\cite{george15}, respectively. Bottom panel: relative differences with respect 
  to the N0 model.
  }
\label{f:zre}
\end{figure}

It is known that the amplitude and shape of the kSZ power spectrum depends on 
many cosmological parameters \citep[see e.~g.][]{shaw12}. With our method, we can easily 
study the dependence on \zre\ by varying the integration limit of our light-cones. We 
show in Fig.~\ref{f:zre} how the amplitude, \dkszt, of the kSZ power spectrum at 
$\ell=3000$ increases as a function of redshift. It is important to note how the kSZ 
effect receives a significant contribution from both low and very high redshift gas: 
again, this is in agreement with previous findings \citep{roncarelli07,shaw12}. More 
precisely, for the N0 model if we assume \zre$=8.8$ (15), then 61 (52) per cent of the 
value of \dkszt\ is due to gas located at $z<2$. When introducing massive neutrinos the 
decrease of the kSZ signal shows no significant differences with redshift, even if the 
damping is slightly more prominent at low redshift where non-linear structure formation 
becomes important. For \smnunz\ and in the range $6<$\zre$<15$ the value of \dksz\ is 
smaller by about 10, 15 and 45 per cent, respectively, with respect to the \lcdm\ model. 
This indicates that the effect of these two variables on the kSZ power spectrum can be 
studied independently. The plot of Fig.~\ref{f:zre} also shows the combined results by 
\cite{planck16cp} on \zre, obtained with the assumption of instantaneous reionization, 
and on \dkszt\ by \cite{george15}: as expected all models fit within 1$\sigma$, 
indicating that more precise measurements of the two observables need to be achieved to 
constrain \smnu.

In order to provide a more precise estimate of the dependence of \dksz\ on \zre\ and 
neutrino mass, we fit our results on the kSZ power spectrum with the following relation:
\begin{equation}
\mathcal{D}^{\rm kSZ}_{\ell} = \mathcal{D}^{\rm kSZ}_{0,\ell} \, 
                               z_{\rm re}^\alpha \, (1-f_\nu)^\beta \, ,
\label{e:fit}
\end{equation}
where $\mathcal{D}^{\rm kSZ}_{0,\ell}$ is the result for our fiducial model that assumes 
\fnu$=0$ (N0 run) and \zre$=8.8$. In Table~\ref{t:fit} we show the best-fitting values 
(last two columns) of $\alpha$ and $\beta$ computed separately for different values of 
$\ell$, together with the \dksz\ value of our N0 run and the same value after applying 
the box volume correction described in Section~\ref{s:sys}. The dependence on \zre\ is 
higher for low multipoles, going from $\alpha=0.48$ at $\ell=1000$ to $\alpha\simeq0.2$ 
at $\ell > 5000$: this trend is justified by the shape of density/velocity perturbations 
at $z\simeq9$ that peak slightly at $5-10$ arcmin.

\begin{table}
\begin{center}
\caption{
  Amplitude of the kSZ effect power spectrum, \dksz, and its dependence with \zre\ and 
  \smnu. First column: multipole $\ell$. Second column: value of \dksz, in \muk\sqr, 
  obtained assuming \smnu$=0$ (N0 run) computed averaging over the 50 light-cones (3.24 
  deg\sqr\ each). Third column: \dksz\ after applying a correction that accounts for a 20 
  per cent of missing signal due to the limited box size (see Section~\ref{s:sys}). The 
  last two columns show the dependence of \dksz\ on the redshift of reionization and 
  neutrino mass fraction, computed as \zre$^\alpha\,\left(1-f_\nu\right)^\beta$ (see 
  equation~\ref{e:fit}).
  }
\begin{tabular}{ccccccc}
\hline
\hline
$\ell$ &&  \multicolumn{2}{c}{\dksz\ $\left(\umu{\rm K}^2\right)$} && 
\multicolumn{2}{c}{\zre$^\alpha \, \left(1-f_\nu\right)^\beta$}  \\
\cline{3-4} \cline{6-7}
       &&  Uncorr. & Vol. corr. &&  $\alpha$ & $\beta$ \\
\hline
 1000 && 1.88 & 2.4 && 0.48 & 12.4 \\
 2000 && 2.79 & 3.5 && 0.29 & 13.6 \\
 3000 && 3.21 & 4.0 && 0.26 & 14.3 \\
 4000 && 3.71 & 4.6 && 0.27 & 15.6 \\
 5000 && 3.74 & 4.7 && 0.24 & 14.2 \\
 6000 && 4.07 & 5.1 && 0.23 & 15.1 \\
 8000 && 4.21 & 5.3 && 0.18 & 15.0 \\
10000 && 4.32 & 5.4 && 0.20 & 15.1 \\
15000 && 4.39 & 5.5 && 0.17 & 16.1 \\
20000 && 4.19 & 5.2 && 0.15 & 16.4 \\
\hline
\hline
\label{t:fit}
\end{tabular}
\end{center}
\end{table}

These results can be compared with the ones by \cite{shaw12}, obtained with a completely 
different approach but similar cosmological parameters: they obtain a steeper dependence 
for both their non-radiative (NR) model, with values of 0.3--0.4, and cooling star 
formation (CSF), with values of 0.4--0.6. This difference with respect to their NR model 
appears to be connected to the treatment of low redshift gas: in fact, their cumulative 
contribution of the kSZ effect (see their fig.~6) shows a fraction of about 50 per cent 
of kSZ effect associated with the redshift range $0<z<2$ of, compared to our 61 per cent. 
This indicates that their method is underestimating the contribution in the non-linear 
regime \citep[for a detailed explanation, see][]{park16}, thus reducing the kSZ at low 
redshift and steepening their relation at higher $z$. In their CSF run this effect adds 
to the overcooling problem present in their simulation that they do not correct (see, 
instead, our correction with equation~\ref{e:rhoc}). This causes a further reduction of 
the low redshift contribution.

The dependence on neutrino mass shows an opposite trend as a function of the multipole, 
increasing from $\beta=14.3$ at $\ell=3000$ to $\beta=16.4$ at $\ell=20000$. 
This is similar to the expected behaviour of the dependence with \sigmae\ 
\citep[see][]{shaw12}, suggesting that even with a measurement of \dksz\ at different 
multipoles it is difficult to break the \smnu--\sigmae\ degeneracy. However it is 
interesting to point out that the dependence with \fnu\ of the tSZ power spectrum is 
significantly different: \cite{roncarelli15} found that its normalization scales as 
\dtsz$\propto\left(1-f_\nu\right)^{25-30}$, so approximately with a double steepness with 
respect to the kSZ equivalent at $\ell=3000$. On the other hand, it is known that the 
dependence of the two power spectra with \sigmae\ shows less differences: in fact, the 
kSZ effect scales as \dksz$\propto$\sigmae$^5$, while the tSZ effect only slightly 
steeper, with \dtsz$\propto$\sigmae$^{7-8}$ \citep{shaw10,shaw12,trac11,horowitz16}. This 
suggests that a combined tSZ-kSZ observation can help break the degeneracy between the 
effect of massive neutrinos and \sigmae. This point can be understood considering 
that, unlike the kSZ, the tSZ effect is dominated by small-scale perturbations of the IGM 
(i.~e. galaxy clusters and groups). At large wave numbers the suppression of the 
amplitude of the matter power spectrum induced by neutrinos is almost scale-independent, 
therefore, at that regime, the effects of massive neutrinos and variations of \sigmae\ 
are degenerate, and observables that are sensitive to these scales will exhibit 
that behaviour. A standard way to break this type of degeneracy consists in combining 
observables on small scales, such as the halo mass function of the Ly-$\alpha$ forest, 
with the ones more sensitive to large scales like CMB, BAO or, as discussed here, kSZ 
effect power spectrum.


\section{SYSTEMATIC EFFECTS} \label{s:sys}
One of the main challenges in modelling the kSZ effect properties with simulations is the 
necessity of runs with both large volumes and high resolution: in fact, the kSZ effect is 
the result of the coupling between the large-scale velocity modes and the small-scale
non-linear gas motions. In this section we will use our second simulation set to 
investigate the possible systematics in our modelling of the kSZ power spectrum due to 
finite resolution and limited box size. In addition, we will use our CSFW run to assess 
the reliability of our physical treatment of the IGM in describing the kSZ properties.

\begin{figure}
\includegraphics[width=0.5\textwidth]{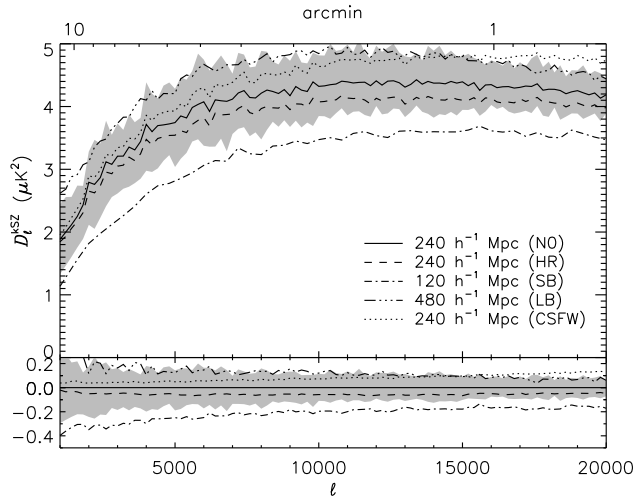}
\caption{
  Same as Fig.~\ref{f:pows} but for the second simulation set, i.~e. assuming massless 
  neutrinos with varying simulation parameters and physics. The solid line shows the 
  average obtained from our reference model, N0, with the grey-shaded area that encloses 
  the central 34 light-cones (68 per cent of the total). The dashed, dot-dashed, 
  dot-dot-dot-dashed and dotted lines correspond to the average of the HR, SB, LB and 
  CSFW simulations, respectively (see details in Table~\ref{t:sim}).
  }
\label{f:pows_syst}
\end{figure}

The plot in Fig.~\ref{f:pows_syst} shows the kSZ effect power spectrum as a function of 
the multipole $\ell$ assuming \zre$=8.8$ and massless neutrinos but varying simulation 
parameters. Since the N0 and HR simulations differ only for the mass resolution, a 
comparison between the two isolates the dependence on this parameter. The two power 
spectra are almost identical at low $\ell$, while the amplitude of the HR simulation is 
lower by about 5 per cent for $\ell > 3000$. This small difference is due to the star 
fraction correction, described by equation~(\ref{e:rhoc}): since the HR simulation 
predicts a significantly higher star formation (see Fig.~\ref{f:fstar}) and since we 
apply the same correction on all the particles belonging to the same redshift and not in 
the exact position where star particles are formed, this perfectly accounts for the 
missing power of the HR simulation at large scales, but fails to recover the full signal 
at small ones. The fact that despite the extreme value of \fstar$(z)$ predicted by the HR 
model we are able to get almost the same result as for the N0 power spectrum, indicates 
that for more realistic star formation histories, like the ones of our first simulation 
set, our method is able to mimic the star-formation history of \cite{ilbert13} without 
introducing significant systematics due to the lack of spatial information. This is 
partially confirmed by the analysis of the CSFW power spectrum, which shows a slightly 
higher amplitude with respect to the N0 one (less than 5 per cent for $\ell < 5000$). 
However at higher multipoles the IGM feedback starts to play a non-negligible role, with 
galactic winds that increase the difference up to 10 per cent at $\ell \simeq 15000$ 
\citep[see also the discussion in][]{roncarelli07}.

\begin{figure}
\includegraphics[width=0.5\textwidth]{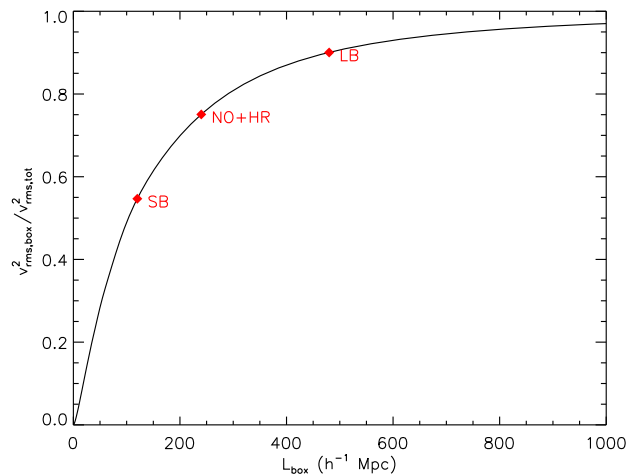}
\caption{
  Fraction velocity power as a function of simulation box size in the linear-theory 
  approximation. The red diamonds show the values of our second simulation set (see 
  Table~\ref{t:sim} for details).
  }
\label{f:linpow}
\end{figure}

We test the impact of limited volume by using the results of the other two 
simulations, SB and LB, and comparing them with an analytical prediction. \cite{iliev07} 
defined a method to estimate the amount of missing velocity power based on the 
linear theory: we refer the reader to their work for the details of the procedure, and 
summarize the final result. When the power spectrum of the velocity field, $P_{vv}(k)$, 
is known it is possible to calculate its rms in the following way 
\begin{equation}
v_{\rm rms}^2 = \int_{k_{\rm min}}^{k_{\rm max}} \! \frac{k^2P_{vv}(k)}{2\upi^2} \, {\rm d}k \, ,
\label{e:vrms}
\end{equation}
where $k$ is the wave number. Since $P_{vv}(k)$ decreases rapidly at small scales, the 
result depends on the value of ${k_{\rm min}}=1/L_{\rm box}$ that can be adjusted 
according to the sampled volume. Therefore we can compute the density power spectrum 
$P_{\delta\delta}(k)$ in the linear-theory approximation using \textsc{camb} with our 
reference \lcdm\ cosmology\footnote{We verified that the result on the missing power in 
the linear-theory approximation does not change significantly when introducing non-zero 
\smnu.}, then derive $P_{vv}(k)$ using the mass conservation equation and compute the 
value of $v_{\rm rms}^2$ as a function of the simulation box size. Fig.~\ref{f:linpow} 
shows the fraction of total linear-theory velocity power as a function of $L_{\rm box}$.

This method predicts that our N0 simulation misses the 25 per cent of the total linear 
velocity power, while about half and 10 per cent should be missed by the SB and LB runs, 
respectively. Obviously the linear prediction serves only as an indication of possible 
biases since at the angular scales of interest for this work the kSZ effect receives a 
significant contribution from the low-redshift Universe (see Fig.~\ref{f:zre}), thus also 
by gas already in the non-linear regime whose dynamics should be well described by our 
hydrodynamical simulations. Analysing the results of Fig.~\ref{f:pows_syst}, we do 
observe a significant damping for the SB with respect to the N0 run: about 20 per cent at 
large $\ell$, with respect to about 30 per cent predicted by the linear theory. Most 
interestingly, the value of \dkszt\ for the LB model is 3.73 \muk\sqr, so the N0 result 
of 3.21 \muk\sqr\ is lower by 14 per cent, with respect to almost 20 per cent predicted 
by the linear theory. This indicates that although our 240\hmone\ Mpc box is lacking part 
of the large-scale velocity power, the linear theory approximation slightly overestimates 
this amount. With the data that we have it is difficult to provide a precise value of the 
amount of kSZ signal that is missing due to the limited box size, and its dependence on 
the angular scale. However, given what said above, we estimate it to be about 20 per cent 
and add the volume corrected values of \dksz\ in the third column of Table~\ref{t:fit}.

When accounting for this correction, we point out that our prediction on the kSZ power 
spectrum at $\ell=3000$ for the \lcdm\ model becomes \dkszt$=4.0$ \muk\sqr, thus 
significantly higher with respect to what obtained by several previous works, such as 
\cite{battaglia10}, \cite{trac11} and \cite{shaw12}, even if we are assuming a lower 
value of \zre$=8.8$ instead of the more popular \zre$=10$. Again, this is associated with 
the fact that previous simulations did not correct for the overcooling problem, hence 
overestimating the amount of star formation that, instead, we calibrate on observational 
data. On the other hand, the estimate of \cite{shaw12} is likely subject to an 
underestimate of the kSZ signal at low redshift (see the discussion in 
Section~\ref{ss:pows}). An important consequence is that when applying this 
correction our constraints on patchy reionization become significantly tighter with 
respect to previous estimates. In detail, combining our result with the SPT measurement 
of \dkszt$=(2.9\pm1.3)$ \muk\sqr\ \citep{george15}, we obtain an upper limit of 
\dksztp$<1.0$ \muk\sqr\ (95 per cent C.~L.).


\section{SUMMARY AND CONCLUSIONS} \label{s:concl}
In this paper we have analysed a set of four cosmological hydrodynamical simulations to 
study how massive neutrinos influence the properties of the kSZ effect due to the IGM 
after reionization. Our runs have been carried out with the \textsc{treepm}+SPH code
\gadgetiii, following the evolution of three different particle types: gas, CDM and 
massive neutrinos in a box of 240\hmone\ Mpc per side. Starting from their outputs at 
different redshifts we have produced 50 light-cone realizations from which we computed a 
set of Doppler \bpar\ maps for each model, and relative kSZ power spectrum keeping track 
of the information in 20 redshift bins in the interval $0<z<15$. Since our simulations 
followed star formation with a simplified model we corrected their output value of 
\fstar$(z)$ in order to match the current observational results \citep[UltraVISTA data 
by][]{ilbert13}. In addition, we have done the same analysis on a set of four simulations 
with massless neutrinos but varying the resolution, box size and physical prescription, 
to study possible systematic effects.

Our results can be summarized as follows.
\begin{enumerate}
\item In agreement with previous findings \citep{roncarelli07}, the typical values of the 
relative temperature variations are of the order of $10^{-6}$, with peaks of $10^{-5}$. 
For all models the distribution of temperature fluctuations is well represented by a 
Gaussian curve, whose dispersion decreases with increasing neutrino mass as 
$\sigma_{b,\nu} = \sigma_{b,0}(1 - f_\nu)^{5.2}$.
\item The kSZ effect at large scales can be described with good precision (about 
5 per cent at $\ell < 5000$) adopting simulations with simple physical prescription, by 
applying a correction that accounts for the uncertainty in the star mass fraction.
\item Using the second simulation set we verified how our results depend on the limited 
box size, and verified that the linear-theory approximation slightly overestimates the 
amount of missing velocity power. Given our box size of 240\hmone\ Mpc per side we 
estimate that we are missing about 20 per cent of the total kSZ power spectrum, instead 
of the 25 per cent predicted by the linear theory.
\item In our \lcdm\ model we obtain a value of \dkszt$=3.21$ \muk\sqr, which becomes 
\dkszt$=4.0$ \muk\sqr\ if we add the expected missing velocity power, enough to account 
for all of the SPT measurement of \dkszt$=(2.9\pm1.3)$ \muk\sqr\ \citep{george15}. With 
currently favoured cosmological parameters, this points towards a rapid reionization 
scenario and translates into a stringent upper limit to the possible kSZ effect 
contribution from patchy reionization models of \dksztp$<1.0$ \muk\sqr\ (95 per cent 
C.~L.).
\item Massive neutrinos induce a reduction of the kSZ power spectrum at all angular 
scales of about 8, 12 and 40 per cent for \smnunz, respectively. All models agree within 
1$\sigma$ with the SPT measurement indicating that current kSZ data alone can not provide 
interesting constraints on \smnu.
\item We studied how the value of the kSZ power spectrum scales with \zre\ and \smnu\ 
(see Table~\ref{t:fit}). At $\ell=3000$, with out fiducial values, we obtain  
\dkszt$\propto$\zre$^{0.26}(1-f_\nu)^{14.3}$. We observe that the dependence on 
$(1-f_\nu)$ is significantly shallower with respect to the tSZ equivalent, that is 
expected to scale as \dtsz$\propto(1-f_\nu)^{25-30}$ \citep{roncarelli15}. Since the 
scaling of the two quantities with \sigmae\ is more similar (\sigmae$^5$ and 
\sigmae$^{7-8}$ for the kSZ and tSZ effect, respectively), this suggests that a combined 
measurement of the two power spectra may represent an additional way to break the 
\smnu--\sigmae\ degeneracy.
\end{enumerate}

Despite the relatively large error bars of the first kSZ power spectrum measurement 
\citep{crawford14,george15}, our results show that this type of observable is extremely 
promising in providing information on the history of LSS formation. In fact, we have 
shown that with current \lcdm\ cosmology the kSZ due to post ionization already exceeds 
the nominal value and is close to the $+1\sigma$ limit, providing interesting constraints 
on several astrophysical and cosmological models. Indeed, besides patchy reionization, 
several non-standard cosmologies, like quintessence \citep[][]{Ratra_Peebles_1988,
Wetterich_1988}, modified gravity \citep[see e.~g.][]{Hu_Sawicki_2007}, and coupled dark 
energy \citep[][]{Wetterich_1995,Amendola_2000,Baldi_2011a}, generically predict an 
excess of velocity power spectrum with respect to standard \lcdm\ 
\citep{Jennings_etal_2012,Li_etal_2012,ma14,bianchini16,Baldi_Villaescusa-Navarro_2016}. 
On the other hand the presence of massive neutrinos, which represent a necessary 
ingredient for any realistic cosmological model, results in a reduction of the kSZ effect 
and must be taken into account before rejecting any of the mentioned model based on their 
predicted excess in the kSZ signal.

In this framework further theoretical analyses are required, in particular simulations of 
larger cosmological volumes to account for the full velocity power spectrum and a more 
extended sampling of the parameter space for both standard (as e.~g. \omegam\ and 
\sigmae) and non-standard (as e.~g. dark energy and modified gravity) cosmological 
parameters aimed at studying possible intrinsic degeneracies. These detailed models, that 
we plan for our future works, will be necessary in the perspective to interpret current 
and possible future kSZ effect observations by e.~g. \planck\, ALMA, ACT and ACTPol or 
the future generation of microwave telescopes.

\section*{Acknowledgements} 
This work has been supported by ASI (Italian Space Agency) through the Contract 
n.~2015-046-R.0. MR also acknowledges financial contribution from the agreement ASI 
n.~I/023/12/0 ``\emph{Attivit\`a relative alla fase B2/C per la missione Euclid}''. The 
work of FVN is supported by the Simons Foundation. The work of FVN has been supported by 
the ERC-StG ``cosmoIGM'' and partially supported by INFN IS PD51 ``INDARK''. 
MB acknowledges support from the Italian Ministry for Education, University and Research 
(MIUR) through the SIR individual grant SIMCODE, project number RBSI14P4IH. We thank an 
anonymous referee that provided useful comments that improved the quality of our work. We 
also thank C.~Carbone, S.~Ettori, L.~Moscardini, E.~Vanzella, M.~Viel and G.~Zamorani for 
useful suggestions and discussions.

\bibliographystyle{mn2e}
\newcommand{\noopsort}[1]{}

\label{lastpage}
\end{document}